\documentclass{elsart}
\usepackage{natbib}

\usepackage{graphicx}

\begin{document}
% running author title
\runauthor{Augusto et al.}
% title, authors, acknowledgments:
\begin{frontmatter}
\title{The kpc-scale radio source population}
\author[Mad]{P. Augusto\thanksref{TMR}}
\author[Cant]{J.I. Gonzalez-Serrano}
\author[Durh]{A.C. Edge} 
\author[Ioa]{N.A.B. Gizani}
\author[Jod]{P.N. Wilkinson}
\author[Can]{I. Perez-Fournon}

\thanks[TMR]{The author acknowledges that this research was
supported by the European Commission's TMR Programme under contract
No.\ EERBFMGECT 950012.}
\address[Mad]{Universidade da Madeira, Centro de Ci\^encias Matem\'aticas,\\
Caminho da Penteada, 9000 Funchal, Portugal}
\address[Cant]{Instituto de F\'{\i}sica de Cantabria (CSIC-Universidad de Cantabria),\\ Facultad de Ciencias, 39005 Santander, Spain} 
\address[Durh]{University of Durham, Dep.\ of Physics, South Road, Durham DH1 3CE, UK} 
\address[Ioa]{University of Ioannina, Section of Astro-Geophysics, Dep.\ of Physics,\\ 45110 Ioannina, Greece} 
\address[Jod]{University of Manchester, Nuffield Radio Astronomy Laboratories, Jodrell Bank, Macclesfield, Cheshire SK11 9DL, UK} 
\address[Can]{Instituto de Astrof\'{\i}sica de Canarias, c/ Via L\'actea s/n, 38200 La Laguna, Tenerife, Spain}

% an abstract (must be included!)
\begin{abstract}
 
   We are conducting a multi-wavelength (radio, optical, and X-ray) 
observational campaign to classify,
morphologically and physically,
a sample of 55 flat-spectrum radio sources dominated by 
structure on kpc-scales. This sample contains  
22 compact-/medium-sized symmetric object 
candidates, a class of objects thought to be the early stages of 
the evolution of radio galaxies. The vast 
majority of the remaining objects have core-plus-one-sided-jet 
structures, half of which 
present sharply bent jets, probably due to strong interactions with 
the  interstellar medium of 
the host gala\-xies. Once 
the observational campaign is completed, we will constrain 
evolutionary theories of 
radio galaxies at their intermediate stages and possibly understand the physics 
 of the hypothesized narrow
line region in active galactic nuclei, given our advantageous statistical position.

\end{abstract}
% enter no more than 6 relevant keywords (see the file keywords.txt
% for a complete list)
\end{frontmatter}
\section{Introduction}
\label{intro}

%\subsection{Motivation and Sample}
%\label{intro.mot}
\citet{Augetal98} conducted the first systematic  search for flat-spectrum 
radio sources with dominant structure on
90--300 mas angular scales (0.2--2 kpc linear scales at $z>0.2$):
 gravitational lenses and compact-/medium-sized
symmetric objects (CSO/MSOs), in particular. 
The selected sample of 55 such radio sources 
is described in Section~\ref{sample}. In
Section~\ref{RGqua} we discuss results, from this sample, 
pertaining to CSO/MSOs. These are
symmetric double or triple sources, with sizes smaller than 15 kpc 
(e.g., \citet{Reaetal96a,Reaetal96b}).
CSOs ($<1$ kpc; aged $10^3$--$10^4$ years) and  MSOs (1--15 kpc; aged $10^5$--$10^6$ years)
present compact lobes ($<20$ mas) having, overall, $\alpha < 0.75$  
($S_{\nu} \propto \nu^{-\alpha}$), and 
are probably the
 precursors of the large radio 
galaxies which they resemble. 
VLBI surveys 
have unveiled a significant population of eighteen 0.01--0.1 kpc CSOs, which constitute $\sim6\%$ of
 a complete flux-limited sample of 293 flat-spectrum  radio sources (CJF; 
\citet{Tayetal96a,Tayetal96b}).
In the last section, 
we mention the potential of our sample in terms of understanding the
 hypothesized  kpc-sized 
narrow-line region (NLR)
in active galactic nuclei (AGN; e.g., \citet{Rob96}).

\section{The Sample}
\label{sample}

Starting from the total of $\sim4800$ sources in the Jodrell-VLA 
Astrometric Survey (e.g.,
\citet{Patetal92}) and
the first part of the Cosmic Lens All Sky Survey (e.g., \citet{Broetal98}), \citet{Augetal98} have first 
established a parent sample containing 1665 
strong ($S_{\rm 8.4\: GHz}>100$ mJy),
flat-spectrum ($\alpha_{1.4}^{4.85}<0.5$) radio sources. 
From this sample, 55 sources were selected in accordance with
an extra resolution criterion as 
described in \citet{Augetal98}. The completeness of this latter sample depends
on both the separation and the flux density ratio of the components of each radio 
source in the parent sample. Unresolved single-component sources would be rejected.

%By imposing resolution on the 
%existing VLA
%8.4 GHz maps ($\sim200$ mas beam), we selected a sample of 55 radio sources. 
%The completeness of this sample  depends on both the separation and flux-density
%ratio of components \citep{Augetal98},
% but
% the extreme cases are: 
%(i)  sources with at least
%two components having a separation of 160--300 mas and a flux-density ratio  
%$\geq$ 1/7; and (ii)  sources with at
%least two roughly equal-brightness components with a separation of 90 mas. 

With regard to the spectral properties of the 55-source sample \citep{Augetal98},
45 sources have power-law radio spectra down to the
lowest measured frequency (which is 365 MHz for 31 of the objects and
151 MHz for 14 of them), 3  sources present complex spectra, and
 7 have spectra peaked at $\sim300$ MHz. It is relevant that only two of the fourteen
0.2--1 kpc CSO candidates in Table~\ref{CSOMSO} can be classified as GHz-Peaked Spectrum Sources (GPSs),
peaking at
$\sim0.5$--10 GHz. From the same table, only three CSO candidates have  a peak at $\sim300$ MHz.
Hence, the statement 
``every CSO is a GPS source'' \citep{Bicetal97} seems incorrect. 
 
There are two main populations of radio sources uncovered on kpc scales by \citet{Augetal98}.
 These consist of 22 CSO/MSOs  and 30 core-plus-one-sided-jet (CJ) 
sources. 

%Each population
%splits roughly  into half. For the CSO/MSOs, there are the ``bright core'' 
%and ``faint core'' 
%populations, as defined in \citet{Augetal98}.
%For the CJs, there are ones with nearly straight jets (with bends
%smaller than $90^{\circ}$) and others with 
%sharply bent jets (bending by more than
%$90^{\circ}$, in some cases more than once). 

 It is unfortunate that, for the vast majority of the
55 sources,  information on any optical counterparts comes only from the Palomar Observatory
Sky Survey (POSS) plates. Using  POSS identifications, we have compared
\citep{Augetal98} the abundance of blue stellar objects, red stellar objects,
galaxies, and empty fields for the 55-source and the parent 1665-source samples. We have
found that the fraction of blue stellar objects in the 55-source sample is half of the 
fraction of such objects in the 1665-source sample. 
Furthermore, the fraction of galaxies is three times
larger in the 55-source sample. For the other two identification types examined, the 
results are comparable in both samples (one-third are empty fields 
and one-eighth are red stellar 
objects).

Thus, it seems that selecting kpc structure 
in the radio leads  to selecting structure in the optical.  There seems to exist a global bias
against radio/optical unresolved sources. As regards 
redshift information, we have $<\!\!z\!\!> \: \sim 0.7$ for the 19 out of the 55 sources
that have spectroscopic data. 
The faintest sources (namely, the 18 sources that correspond to POSS
empty fields) still need redshift determinations, suggesting that the average
redshift of the sample will increase. Unfortunately for the discussion on this paper, very
few of the CSO/MSOs have measured redshifts (Table~\ref{CSOMSO}).

 %The comparison of the radio structures of the sample of 55
%sources with the POSS optical identifications gives two
%significant results \citep{Augetal98}: $\sim90\%$ of the  blue stellar objects are
%CJs (QSO candidates) and about $\sim75\%$ of the CSO/MSO candidates are either  galaxies or
% EFs. If the typical host galaxy of CSO/MSO is a $V\sim22$ mag elliptical 
%(e.g.\ Readhead et al., 1996a), then we may interpret these EFs as galaxies below 
%the POSS detection limit.

%\section{Radio galaxies, quasars, AGN and kpc-scales}
%\label{RGqua}

\section{Compact/Medium Symmetric Objects}
\label{RGqua}

The sample of 22 CSO/MSO candidates found by \citet{Augetal98} --- Table~\ref{CSOMSO} ---
 contains 9 certain CSOs and two certain
MSOs. Most likely, six of the remaining sources are MSOs, leaving five sources that could be either.
The fact that for sources at $z>0.2$ we have selected the ones dominated by structures on
0.2--2 kpc scales suggests a bias against the population of MSOs within our sample. Since most
flat spectral-index sources will consist of a pair of compact lobes (plus,
possibly, a core), they will be included in our sample only if their sizes are $\leq0\rlap{.}''3$.
Much larger sources (like B0824+355 in Table~\ref{CSOMSO}), consisting of very weak jets and low
surface brightness extended lobes, are probably the exception. We believe that most MSOs in our
sample will have sizes of $\sim$1--2 kpc, much like the confirmed MSO B0205+722 (Table~\ref{CSOMSO}).
Note that even if we allow for sources with $z<0.2$, this will only favour the increase
in number of small MSOs, since for the same angular dimensions seen, a lower redshift
will translate into a smaller
linear size.

%The bulk of the 55-source sample cannot tell much about MSOs, since only the smallest ones
%(1--2 kpc) are likely to be included. Nevertheless, there ought to be some large MSOs in the
%sample (e.g., 0824+355).

\citet{Augetal98} have shown that the 55-source sample includes every CSO from the 1665-source
parent sample having a 160--300 mas separation (0.4--1 kpc for sources
at $z>0.2$) between compact components with a flux-density ratio of
7:1 or smaller.
CSOs containing compact lobes with similar
 flux-density
ratios are included in the sample down to a separation of 90 mas (0.2--0.6 kpc at $z>0.2$).
The key issue now is to review evidence for why virtually all CSOs present
in the 1665-source parent sample are at most the 14 found by \citet {Augetal98}
among their 55-source sample.
Typically,
CSOs have weak cores (weaker than any of the lobes) and, hence, it is the
lobes that are the `components'
that will go through the selection criterion of \citet{Augetal98}.  It is
very rare to find a CSO with lobes presenting flux density ratios greater than 
7. In fact, there are
not any of these cases among the 0.2--1 kpc CSOs in Table~\ref{CSOMSO} or the eighteen 0.01--0.1 kpc CJF 
CSOs discussed here. Therefore, we believe
that \citet{Augetal98} have selected virtually all of the CSOs present in the  1665-source parent sample;
these are shown in Table~\ref{CSOMSO}.
In any case, for the discussion of this paper, we performed simulations (see below), which
estimate the effects of the `resolution' criterion on the CJF sample, before 
making any comparison between our CSO-fraction and that of the CJF. 
The simulations give results that are consistent with the `typical'
morphology of CSOs just presented.

Conservatively, taking a maximal number of 0.2--1 kpc CSOs as 14
(Table~\ref{CSOMSO}, including the sources classified as `question marks')
out of a parent sample containing 1665 sources,
only $\sim0.8\%$ of flux-limited samples seem to be such CSOs. It seems, then, that these 
are six times less common than 0.01--0.1 kpc CSOs (which constitute $\sim6\%$ of CJF).
Both the 0.2--1 kpc  and the 0.01--0.1 kpc CSOs are dominated 
by components $<20$ mas in size. Is the number difference
 due to luminosity
 evolution alone? Strong luminosity evolution takes place during the time that the
 0.01-kpc scale CSOs grow to be 100-kpc scale radio sources 
(e.g., \citet{Reaetal96a,Reaetal96b}).
\citet{KaiAle97} have  proposed a model in which
 the luminosity of double sources decreases proportionally to the square root of their size. 
If this relation applies continuously as the source evolves 
from the 0.01-kpc to the 100-kpc scale, then 
in the evolution from a 
0.01--0.1 kpc to a 0.2--1 kpc CSO, size increases by a factor of $\sim10$
and hence the luminosity decreases
by $\sim3$. Given that all CJF sources have $S_{\rm 5 \: GHz} > 350$ mJy and, like the 55-source sample,  
have $\alpha_{1.4}^{4.85}<0.5$, our flux-density 
criterion $S_{\rm 8.4 \: GHz} > 100$ mJy allows a sampling
$\sim3$ times fainter, cancelling out the predicted luminosity evolution.
Before rushing to other evolutionary explanations, 
we note that our selection process
 included a resolution criterion not present in CJF. Hence, we need to find out 
how many of the 18
CJF  CSOs   would remain in the CJF if it
had an equivalent resolution criterion. 
 The simplest way to do this is to use models
fitted to the 18 CJF CSOs, expand the separation of the components by
a factor of 10, and check whether they  meet the criteria for inclusion in our sample.
This will only give indicative results, of course.
The models and maps are found in the literature from the VLBI surveys,
except for three models that we crudely produced from the available maps. 

%Being purist, we have
%converted the 1.6 GHz and 5 GHz flux densities to 8.4 GHz flux densities by using the overall
%fluxes of the sources (Patnaik et al., 1992 and model/maps references) and assuming all components
%with the same $\alpha_{1.6}^{8.4}$ and $\alpha_{5}^{8.4}$. 

Using  the program {\sc FAKE} in the Caltech VLBI package \citep{Pea91}, 
we have performed a test for the reliability of selection (details in 
\citet{Augetal98}).
 Eleven out of the `order-of-magnitude-expanded' 18 CJF 0.01--0.1 kpc CSOs would  be in our sample.
To contemplate the possibility that some of our 0.2--1 kpc CSOs might have been selected by a lucky
combination of observational conditions, we also ran {\sc FAKE}  on the 14 such  CSOs in Table~\ref{CSOMSO}.
All of them are reliably in our sample.

The revised frequencies of CSOs are then $\sim0.8\%$ (14/1665) in our sample and $\sim4\%$ (11/293)
in CJF. Since  five of the fourteen  0.2--1 kpc CSO candidates in Table~\ref{CSOMSO} could 
be $>$1 kpc MSOs, a conservative factor of $\sim5$ still remains between
the abundance of CSOs in both samples. To explain this difference, we suggest evolution
 of the lobes in CSOs as they grow --- self-similar growth of radio galaxies: the lobes 
start off as compact hot
spots when 0.01--0.1 kpc apart and expand  until they grow $\sim100$ kpc apart,
as in normal radio galaxies. The number of 0.2--1 kpc young radio galaxies seems to be less
 than the number with sizes 0.01--0.1 kpc due to
 the resolution criterion used to select the 55-source sample
in \citet{Augetal98}: only double (or triple) sources with compact ($<20$ mas) components
are in the sample. The extended lobes of  Compact Steep Spectrum 
($\alpha>0.5$) radio sources, the dominant radio sources on 0.2--1 kpc scales,
cannot be selected by the resolution
criterion of \citet{Augetal98}.

\begin{table}[ht] 
\begin{center}
\caption{The 22 compact/medium symmetric object (CSO/MSO) candidates found in the 55-source 
 sample of \citet{Augetal98}.
 The linear size 
is calculated using $H_{0}$=75 km s$^{-1}$ Mpc$^{-1}$ and $q_0=0.5$. For the vast majority 
of sources, without
 redshift information, formal classification is not possible. Nevertheless, independent
of redshift, any object smaller than 175 mas is a CSO. Furthermore, assuming $z>0.2$
 for the remaining  objects,
 angular sizes larger than 350 mas identify MSOs. Palomar Observatory Sky Survey (POSS)
identifications are with galaxies (G), empty fields (EF), and blue or red stellar objects (BSO,RSO).}
\label{CSOMSO}
\vspace{0.5cm}
\begin{tabular}{lrcccc} \hline \hline
\multicolumn{1}{c}{Source} &  \multicolumn{1}{c}{Angular size} & Linear size & 
Classification & POSS id.\ & z \\
\multicolumn{1}{c}{B1950.0} &  \multicolumn{1}{c}{(mas)} & (kpc) & &  & \\ \hline 
0046+316 &  300 & 0.09 & CSO  & G; 15$^m$ & 0.015 \\
0112+518 &  650 & & (MSO) & EF & \\
0116+319 &  75 & 0.08 & CSO  & G; 16$^m$ & 0.0592  \\
0205+722 &  600 & 3 & MSO  & G; 18$^m$ & 0.895 \\
0225+187 &  225 &  & ? & EF & \\
0233+434 &  120 &  & CSO & EF & \\
0352+825 &  44 &   & CSO & G; 15.5$^m$ & \\
0638+357 &  400 &  & (MSO) & EF & \\
0732+237 &  175 &  & CSO & EF & \\
0817+710 &  225 &  & ? & EF & \\
0819+082 &  275 &  & ? & RSO; 19.3$^m$ & \\
0824+355 &  2000 & 11 & MSO  & RSO; 19.6$^m$ & 2.249 \\
1010+287 &  75 &  & CSO & EF & \\
1058+245 &  900 &  & (MSO) & EF & \\
1212+177 &  100 &  & CSO & RSO; 20$^m$ & \\
1233+539 &  240 &  & ? & BSO; 19$^m$ & \\
1504+105 &  110 &  &  CSO & ?; 16$^m$ & \\
1628+216 &  800 &  & (MSO) & ? & \\
1801+036 &  1200 &  & (MSO) & G; 17$^m$ & \\
1928+681 &  120 &  & CSO & BSO; 20.5$^m$  & \\
1947+677 &  500 &  & (MSO) & EF & \\
2345+113 &  275 &  & ? & G; 19$^m$ & \\ \hline
\end{tabular}
%\end{footnotesize}
\end{center}
%\end{minipage}
%\caption{
%\label{CSOMSO}
\end{table}

\section{Future} 
\label{RGqua.NLR}

Once redshifts are determined for the remaining 36 of the 55 sources, 
we will
not only classify CSO/MSOs correctly, according to their sizes, but also determine the linear 
(projected) sizes of the CJs. 
Most of these CJs might also show evidence for strong shocks in the NLR.
Half of the CJs in the 55-source sample contain sharply bent jets that bend by more than $90^{\circ}$,
in some cases more than once. This
hints at strong interactions
with the interstellar medium of the host galaxies.
Altogether, the CSOs, MSOs, and CJs in our sample will give us clues about 
the  composition
and density of the NLR in galaxies because of their interactions with the 
NLRs of
their hosts. Due to our good statistics, this might  be a useful
step forward towards understanding the standard model of AGN as a whole, 
and the NLR in particular.

\end{document}